\documentclass[a4paper]{PoS}
\usepackage{amsmath,amssymb,amsfonts,ulem,graphicx,footmisc,soul,enumitem,bbm,nicefrac,float,booktabs,array,mwe,dsfont,bm,citesort}
\usepackage[utf8]{inputenc}
\usepackage[dvipsnames]{xcolor}
\newcommand{\barr}[1]{\bar{\bar{#1}}}

\title{3-body quantization condition in a unitary formalism}

\ShortTitle{3-body quantization condition}

\author{\speaker{Maxim Mai}\thanks{The speaker thanks the organizers of the conference for the opportunity to give this talk. This work is supported by the German Research Association (MA 7156/1) and National Science Foundation (grant no. PHY-1452055).}\\
        Institute for Nuclear Studies and Department of Physics, The George Washington University, Washington, DC 20052, USA\\
        E-mail: \email{maximmai@gwu.edu}}

\abstract{
Unitarity identifies all power-law finite-volume effects and is, therefore, the crucial S-matrix principle for a mapping between experimental results and those of Lattice QCD calculations. In this contribution we review how 3-body unitarity constrains the form of the 3-body scattering amplitude parametrized by the tower of isobars. The result is discretized and projected to the irreducible representations of the cubic group, leading to a fully relativistic 3-body quantization condition. The latter is used to deduce the finite-volume excited level spectrum of the $\pi^+\pi^+\pi^+$ system, which agrees nicely with the available lattice results by the NPLQCD collaboration.}

\FullConference{The 36th Annual International Symposium on Lattice Field Theory - LATTICE2018\\
		22-28 July, 2018\\
		Michigan State University, East Lansing, Michigan, USA.}

\begin{document}

\paragraph{Introduction}

Many unsolved questions of QCD involve systems with 3 hadrons. One of the most prominent examples is the so-called Roper-puzzle, which addresses the reversed mass-pattern of the $N(1535)\nicefrac{1}{2}^{-}$ and the excited state of the nucleon $N(1440)\nicefrac{1}{2}^{+}$  compared to the expectations from the constituent quark model. The properties of the first can be accessed from the meson-baryon scattering amplitude, e.g., through a manifestly covariant and unitary approach~\cite{Bruns:2010sv}. Unfortunately, a similar treatment of the positive-parity state, the $N(1440)\nicefrac{1}{2}^{+}$, is more intricate due to its large branching ratio to the $\pi\pi N$ channels, see Ref.~\cite{AlvarezRuso:2010xr} for a recent review of several theoretical approaches. In the meson sector, the spin-exotics are states with quantum numbers which cannot be formed by a quark-antiquark pair. Thus, the more complex structure of these states is believed to be an indicator for the gluonic degrees of freedom in QCD. Many of such hypothetical states cannot decay into two, but only into three pions. This also applies to ordinary mesons such as the  $a_1(1260)$-meson. Thus, the identification of these novel states requires an in-depth understanding of the $\pi\pi\pi$ system forming ordinary excited states.

The only systematical, non-perturbative approach to the properties of strongly interacting systems is Lattice QCD (LQCD). Such numerical calculations are performed on discretized Euclidean space-time in finite volume and (depending on the technical intricacy) at unphysical quark masses. In the quest for time-independent quantities, such as resonance parameters, in systems with three hadrons the non-trivial issues are the finite-volume effects as well the subsequent chiral extrapolation. Therefore, while substantial computational and algorithmic advances have been made over the last years in extracting finite-volume spectra from ab-initio LQCD calculations~\cite{Lang:2014tia,Lang:2016hnn,Kiratidis:2016hda,Woss:2018irj,Beane:2007es,Detmold:2008fn}, their comparison to phenomenology requires a so-called 3-body quantization condition, similar to the well-established L\"uscher's method~\cite{Luscher:1986pf,Luscher:1990ux} in the 2-body sector. Over the last years many explorations have been performed to this end~\cite{Hammer:2017kms,Hammer:2017uqm,Briceno:2017tce,Doring:2018xxx,Sharpe:2017jej,Guo:2016fgl,Hansen:2016ync,Hansen:2016fzj,Hansen:2015zga,Meissner:2014dea,Briceno:2012rv,Bour:2012hn,Kreuzer:2012sr,Polejaeva:2012ut} including alternative techniques~\cite{Agadjanov:2016mao,Hansen:2017mnd} proposed to obtain essential information on the system without the need of explicit parametrization of every reaction channel.

Over the last two years a convenient version of a fully relativistic 3-body quantization condition has been derived from the so-called isobar parametrization of the unitary 3-body scattering amplitude~\cite{Mai:2017vot} using discretization~\cite{Mai:2017bge} and projection of the latter to the irreducible representations of the cubic group~\cite{Doring:2018xxx}. In the present contribution we show the results derived in these works as well as the recent application of this approach in the first-ever calculation of excited (i.e., above-threshold) energy eigenvalues~\cite{Mai:2018djl} of a physical ($\pi^+\pi^+\pi^+$) system. The ground state finite-volume energy for this system has been calculated by the NPLQCD collaboration~\cite{Beane:2007es,Detmold:2008fn}. 

\paragraph{Three-body dynamics}

A relativistic, infinite volume 3-body scattering amplitude in the isobar formalism~\cite{Mai:2017vot} can be expressed in terms of on-shell, 2-body unitary $2\to 2$ amplitudes plus real-valued genuine 3-body interactions. The ``isobar'' refers to the parametrization of the ${2\to 2}$ amplitudes in terms of a dressed $s$-channel propagator with dissociation vertices attached to both ends. As further discussed in Ref.~\cite{Mai:2017vot}, the isobar can be associated with bound states, one or more resonances, or a non-resonant 2-particle amplitude.  The isobar formulation is not an approximation but a re-parametrization of the full 2-body amplitude as shown in Ref.~\cite{Bedaque:1999vb} and also discussed in Ref.~\cite{Hammer:2017kms}. In the following we collect only the main results of the derivation and refer the reader for details to Refs.~\cite{Mai:2017vot,Mai:2017wdv}.

The interaction of three spin-less particles of mass $m$ and out- and in-going four-momenta $q_1,\,q_2,\,q_3$ and $p_1,\,p_2,\,p_3$, respectively, is fully described by the S-matrix ($\mathcal{S}_3$) related to the T-matrix ($\mathcal{T}_3$) via ${\mathcal{S}_3=:\mathbbm{1}+i(2\pi)^4 \delta^4\!\left(\sum_{i=1}^3(q_{i}-p_{i})\right) \mathcal{T}_3}$. In the case of $3\to 3$ scattering the latter consists of a fully connected and a once-disconnected piece, related to the isobar-spectator scattering amplitude $\hat T_{\rm is}$ and isobar propagator $\hat\tau$ as
\begin{align}
\label{eq:t33full}
\langle q_1,q_2,q_3|
\mathcal{T}_3| p_1,p_2,p_3\rangle
&=
\frac{1}{3!}\sum_{n=1}^3\sum_{m=1}^3\,
v(q_{\bar{n}},q_{\barr{n}})
\hat T(q_n,p_m;W)
v(p_{\bar{m}},p_{\barr{m}})\,,\\
\hat T(q_n,p_m;W)
&=
\Bigg(
\hat \tau(\sigma_{\bm{q}_n})\,
\hat T_{\rm is}(q_n,p_m;W)\,
\hat \tau(\sigma_{\bm{p}_m})\,
-2E_{\bm{q}_n}\hat\tau(\sigma_{\bm{q}_n})(2\pi)^3\delta^3(\bm{q}_n-\bm{p}_m)
\Bigg)\,,\nonumber
\end{align}
where $P$ is the total four-momentum of the system, $W^2=P^2$ and $E_{\bm p}=\sqrt{{\bm p}^2+m^2}$. All four-momenta $p_1,\,q_1,...$ are on-mass-shell, and the square of the invariant mass of the isobar reads ${\sigma_{\bm{q}}=W^2+m^2-2WE_{\bm{q}}}$ for the spectator momentum $q$. We work in the total center-of-mass frame where ${\bm P}=\bm{0}$. The dissociation vertex $v(p,q)$ of the isobar decaying in asymptotically stable particles, e.g., $\rho(p+q)\to \pi(p)\pi(q)$, is chosen to be cut-free in the physical energy region, which is always possible. For the present study we choose the dissociation vertex to be of a particularly simple form, $v(p,q):=\lambda f((p-q)^2)$ with $f$ such that it is 1 for $(p-q)^2=0$ and decreasing sufficiently fast for large momentum difference to regularize integrals of the scattering equation. The specific form of this form factor will be given below.

\begin{figure}[t]
\includegraphics*[width=1\linewidth, trim=0 .5cm 0 0]{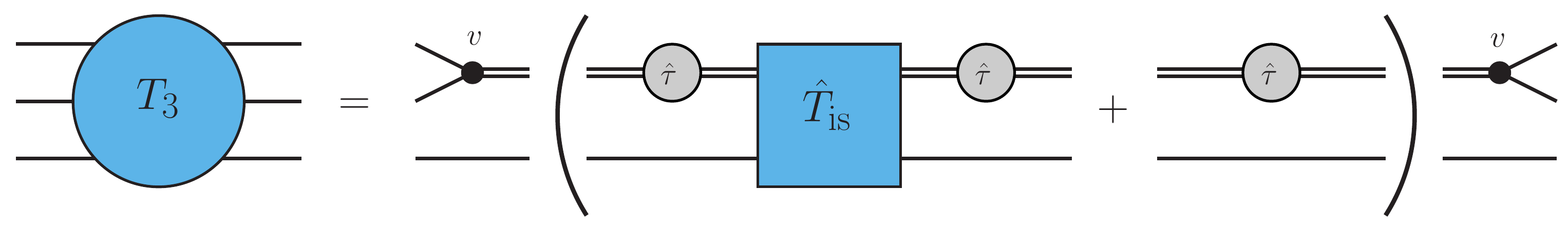} 
\caption{
The 3-particle scattering amplitude $T_3$, constructed from
the particle-isobar scattering amplitude $\hat T_{\rm is}$, isobar propagator $\hat \tau$  and dissociation vertex $v$. The quantity in the parentheses on the right hand side consists of a fully connected and once disconnected parts in that order.
\label{fig:T33}
}
\end{figure}

Imposing 3-body unitarity and a general ansatz for the isobar-spectator scattering amplitude $T$ in Eq.~\eqref{eq:t33full} one obtains
\begin{align}\label{eq:T}
\hat T_{\rm is}(q,p;W)&= 
B(q,p;W)-
\int
\frac{\mathrm{d}^3\bm{l}}{(2\pi)^3}
B(q,l;W)
\frac{\tau(\sigma_{\bm l})}{2E_{\bm l}}
\hat T_{\rm is}(l,p;W)\,,\\\nonumber
B(q,p;W)&=\frac{-\lambda^2f((P-2p-q)^2)f((P-p-2q)^2)}
{2E_{\bm{q+p}}\left(W-E_{\bm q}-E_{\bm p}-E_{\bm{ q+p}}+i\epsilon\right)}
+C(q,p;W)\,,
\end{align}
where $p$ and $q$ denote the on-shell four-momenta of the in- and outgoing spectator, respectively. Additional terms $C$ that are real functions of energy $W$ and momenta in the physical region as demanded by 3-body unitarity (3-body forces) can be added to $B$, see discussion in Ref.~\cite{Mai:2017vot}. We postpone the introduction of multiple isobars and of spin and isospin for the isobars and the stable particles to future work. As demonstrated in Ref.~\cite{Mai:2017vot} the algebraic form of the isobar propagator is fixed up to regular terms and can be written as
\begin{align}
\label{eq:tau}
\frac{1}{\hat\tau(\sigma)}&=
\sigma-M_0^2-
\sum_{\pm}\int \frac{d^3\bm{k}}{(2\pi)^3} 
\frac{(\lambda f((\sqrt{\sigma}\pm 2E_{\bm k})^2-4\bm{k}^2))^2}
{4E_{\bm k}\sqrt{\sigma}(\sqrt{\sigma}\pm2E_{\bm k})}\,,
\end{align}
where $M_0$ is a free parameter that can be used to fit (together with $\lambda$) the 2-body amplitude corresponding to the considered isobar. We will refer to the integral term in Eq.~\eqref{eq:tau} as self-energy in the following. Note that the only principle used in the construction of the amplitude is 3-body unitarity for the physical ($s$-channel) region which is the only requirement needed to identify all power-law finite volume effects; like in the 2-body case, there are, of course, also left-hand singularities in the infinite volume amplitude but they all contribute exponentially suppressed to the finite-volume effects.

\paragraph{Three-body quantization condition}

For the extraction of scattering information from lattice calculations, boundary conditions have to be imposed, and only discrete momenta are allowed. In particular, for a box of side length $L$ and periodic boundary conditions the set of allowed momenta reads $2\pi/L\cdot\mathds{Z}^3$. For convenience, we order these momenta in ``shells'', defined as sets of momenta which are related to each other by cubic symmetry. The running index of these sets will be denoted in the following by $s$ and its cardinality by $\vartheta(s)$.

For discretized momenta, the isobar-spectator amplitude becomes a matrix equation, which in operator notation reads ${T=(\tau^{-1}+B)^{-1}}$ with $\tau$ and $T$ being the isobar-propagator and the full isobar-spectator scattering amplitude (combined terms in parenthesis in Fig.~\ref{fig:T33}) in finite volume, respectively. Obviously, in this symbolic notation $T$ can become singular for 3-body energies $W$ fulfilling ${{\rm det}(\tau^{-1}+B)=0}$. The latter is commonly referred to as the quantization condition and determines the finite-volume spectrum of the 3-body system in question. Note that intermediate states with more than three particles are explicitly excluded from the formalism. This limits the range of validity of the present approach to the next-higher multi-particle channel (being, e.g., $5\pi$ in the case of the $\pi\pi\pi$ system).

Similarly to the infinite volume case, the technical obstacle solving the above (symbolically defined) quantization condition is its high dimensionality. Projection to the irreducible representations (irreps) of the cubic group $\Gamma\in\{A_1,A_2,E,T_1,T_2\}$ reduces the quantization condition greatly and has the advantage that the obtained spectra can be compared directly with the LQCD results. A particularly convenient projection procedure has been introduced in Ref.~\cite{Doring:2018xxx}. Having large similarities to the partial wave projection techniques in infinite volume, it defines an orthonormal basis of functions on each shell. Projection of the above quantization condition to such basis functions is described in detail in Sec.~IV of Ref.~\cite{Doring:2018xxx}, and leads to a diagonal condition in the irrep-index $\Gamma$
\begin{align}
\label{eq:QC3}
\det\left(B^{\Gamma ss'}_{uu'}(W) +\frac{2E_s \,L^3}{\vartheta(s)}
\tau_s^{-1}(W)\delta_{ss'}\delta_{uu'}\right)=0\,.
\end{align}
Here, the determinant is taken with respect to the shell-index $s{}^{(}{'}{}^{)}$ and basis-index $u{}^{(}{'}{}^{)}$, while $E_s:=E_{\bm p}$ and $\tau_s=\tau_{\bm p}$ with ${\bm p}$ being a momentum on the shell $s$. An important consequence of the breakdown of the spherical and therefore Lorentz symmetry is that the isobar-propagator has to be boosted into the isobar rest frame before discretizing the momenta. Denoting the boost of the momentum $\bm{x}$ by the momentum $\bm{p}$ (spectator momentum) as $\bm k^*_{\bm{x,p}}$ and the corresponding Jacobian by $J_{\bm{p}}$, the isobar propagator in finite volume reads
\begin{align}
\label{eq:tauFV}
\frac{1}{\tau_{\bm{p}}}
&=\sigma_{\bm p}-M_0^2
-\frac{J_{\bm{p}}}{L^3}
\sum_{\bm{x}\in \frac{2\pi}{L}\,\mathds{Z}^3}
\sum_\pm
\frac{
\left(
\lambda
f\left(
   \left(
      P^*_{\bm{p}}\pm2k^*_{\bm{x},\bm{p}}
   \right)^2
 \right)
\right)^2}
{4 \sqrt{\sigma_{\bm{p}}}
E_{\bm{k}^*_{\bm{x},\bm{p}}}
\left(
\sqrt{\sigma_{\bm{p}}}\pm
2E_{\bm{k}^*_{\bm{x},\bm{p}}}
\right)}\,.
\end{align}
Here $P^*_{\bm{q}}:=(\sqrt{\sigma_{\bm{q}}},\bm{0})$ is the four-momentum of the isobar (2-pion system) boosted to its reference frame. For a given absolute value of the spectator momentum $\bm{p}$, the range of validity of the boost formula and, therefore, of the discretized propagator $\tau$ is limited to $\sigma_{\bm p}>0$. However, already below the 2-particle threshold $\sigma_{\bm p}<(2M)^2$ the regular summation theorem applies and the sum can be replaced by the integral up to exponentially suppressed terms.

In conclusion, we note that both the isobar-spectator kernel $B$ as well as $\tau^{-1}$ in Eq.~\eqref{eq:QC3} can become singular separately. However, these singularities cancel each other exactly as shown explicitly in Ref.~\cite{Mai:2017bge}, leaving one with the singularities from genuine three-body dynamics only. 

\paragraph{Finite-volume spectrum of the $\pi^+\pi^+\pi^+$ system}

\begin{figure}[t]
\begin{center}
\hspace{1cm} $\bm{\pi}^+\bm{\pi}^+$ \hspace{6.5cm} $\bm{\pi}^+\bm{\pi}^+\bm{\pi}^+$
\\[-25pt]~
\end{center}
  \includegraphics[width=0.49\linewidth]{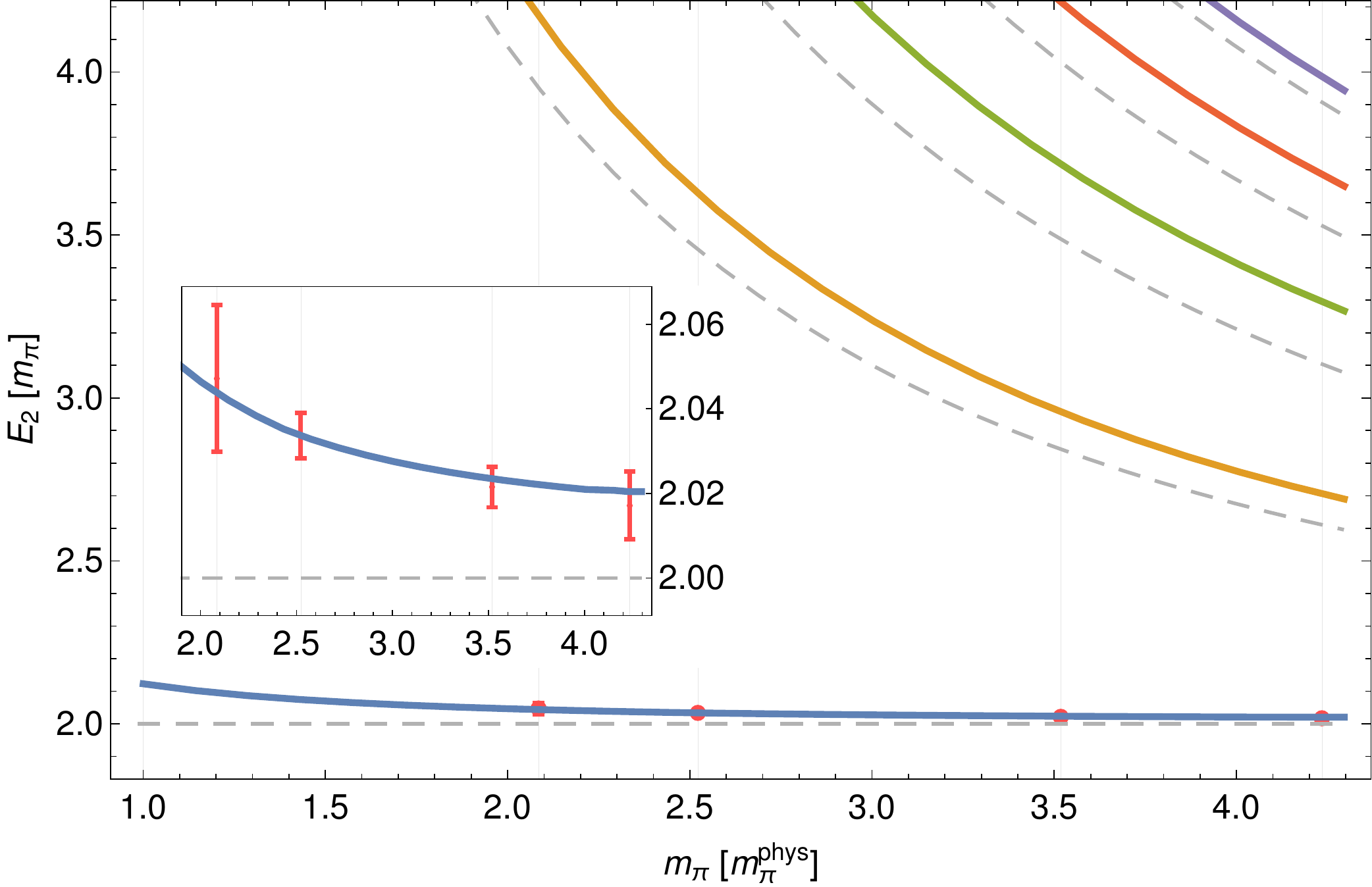}
  ~
  \includegraphics[width=0.49\linewidth]{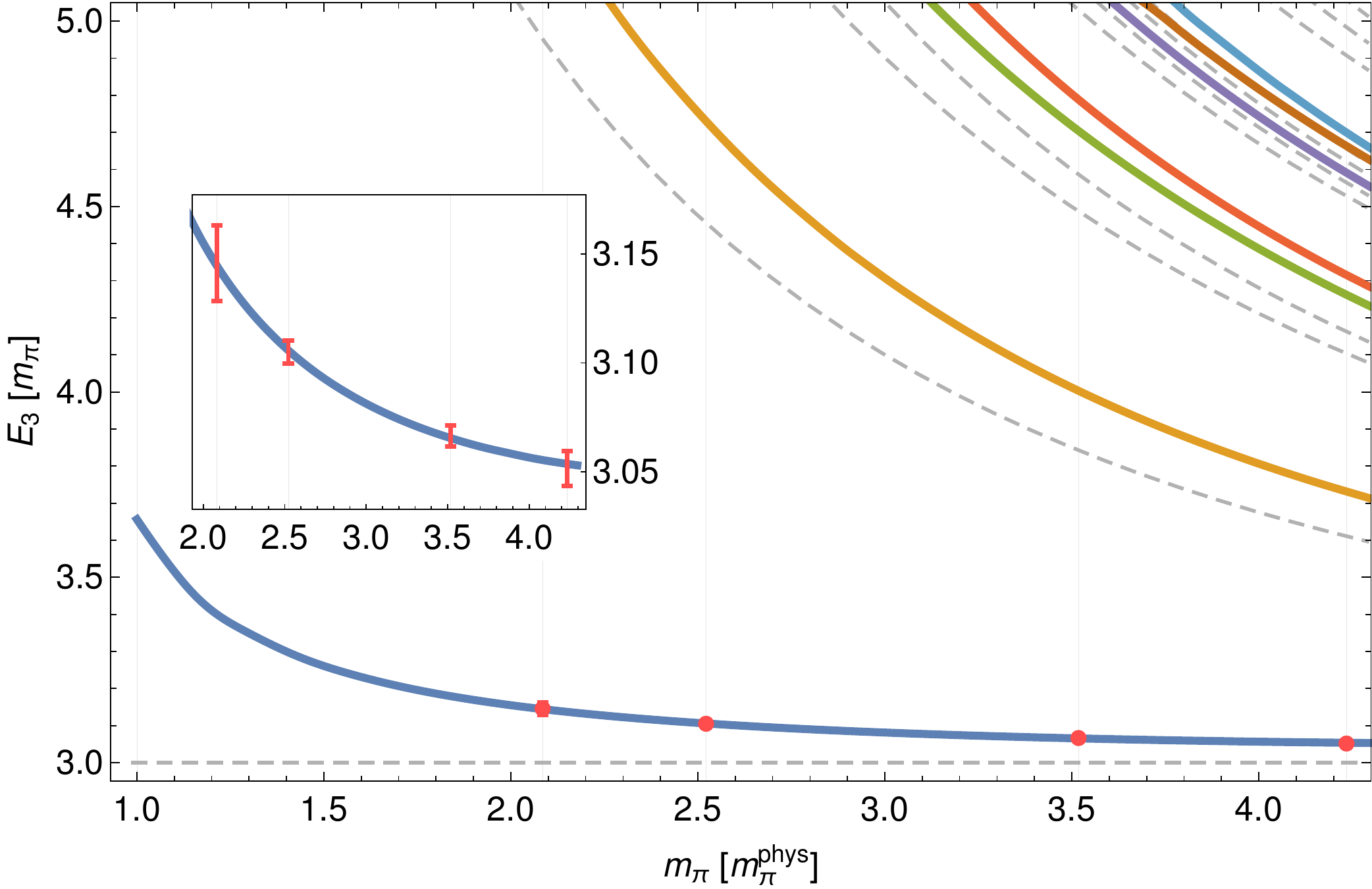}
\caption{
\textbf{Left}: Prediction of 2-body energy levels (full) as a function of $m_\pi$ with dashed lines denoting non-interacting levels. \textbf{Right}: Prediction of excited energy levels for the $\pi^+\pi^+\pi^+$-system as a function of pion mass with non-interacting levels represented by dashed lines. The insets show the zoom-in on the ground level, where the lattice data~\cite{Beane:2007es,Detmold:2008fn} are shown in red.}
\label{pic:2body}
\end{figure}

The quantization condition derived in Eq.~\eqref{eq:QC3} has a particularly simple form and has been tested on a hypothetical scenario of three spin-less particles, two of each interacting via a Breit-Wigner like resonance, see Ref.~\cite{Mai:2017bge}. However, no assumptions have been made in the derivation of the quantization condition about the form of the 2-particle interaction in the sub-channels. In the following we demonstrate the application of the quantization condition~\eqref{eq:QC3} to the physical system of $\pi^+\pi^+\pi^+$. The 2-body sub-channel interaction of this system is repulsive and serves, therefore, as an ideal test bed for the applicability of the proposed 3-body quantization condition. Fortunately,  LQCD data in this (repulsive) channel are available from the NPLQCD Collaboration~\cite{Beane:2007es,Detmold:2008fn} for $L=2.5$~fm and $m_\pi\in\{291,352,491,591\}$~MeV. Our program consists of prediction of the full (up to the $4\pi$ threshold) finite-volume 2-body spectrum using experimentally available data. Subsequently, we will fix the remaining parameter (genuine 3-body coupling) to the ground-state energy level of the $\pi^+\pi^+\pi^+$ system~\cite{Beane:2007es,Detmold:2008fn}, predicting higher levels up to $W=5m_\pi$.

In the following we specify the parameters of the quantization condition~\eqref{eq:QC3} following the findings of Ref.~\cite{Mai:2018djl}. First, the system in question is in relative $S$-wave such that for the finite-volume analysis we fix $\Gamma=A_1^+$. The form-factor $f(Q^2)$ ($Q$ being the difference of the four-momenta of the dissociation products) yields a smooth cutoff of an otherwise log-divergent self-energy part of the isobar propagator (third term in $\tau^{-1}$ of Eq.~\eqref{eq:tauFV}). Note, that this cutoff-dependence cancels in the full quantization condition~\eqref{eq:QC3} by the functions $C$ and $M_0$. Specifically, we chose here ${f(Q^2)=1/(1+e^{-(\Lambda/2-1)^2+ Q^2/4})}$ with $\Lambda=42$ in units of $m_\pi$. Second, we have tested various forms of the coupling $\lambda$ and found that taking 
\begin{align}
\lambda^2=(M_0^2-\sigma)\left(\frac{d}{4\pi^2}+\frac{T_{\rm LO}-\bar T_{\rm NLO}}{T_{\rm LO}^2}\right)^{-1}\,,
\end{align}
where $T_{\rm LO}$ and $\bar T_{\rm NLO}$ are the leading and next-to-leading (without the s-channel loop) order chiral amplitudes~\cite{Gasser:1983yg}, respectively, yields the Inverse Amplitude Method (IAM) for $T_{2}:=v\hat\tau v$. Here $d=0.86$ for the chosen $\Lambda$, see Ref.~\cite{Mai:2018djl} for more details. Such an amplitude has a substantially larger range of validity in the 2-body energy and resembles the chiral expansion up to next-to-leading order exactly as argued in Ref.~\cite{Truong:1988zp}. Indeed, taking the low-energy constants from Ref.~\cite{Gasser:1983yg} we have found that this ansatz perfectly reproduced the phase shifts from experiment. The predicted finite-volume spectrum of the $\pi^+\pi^+$ system, extracted from the corresponding equation when replacing $\hat\tau\mapsto\tau$ lies on top of the LQCD data~\cite{Detmold:2008fn,Beane:2007es} as depicted in Fig.~\ref{pic:2body}. This approach agrees with the lattice data even at pion masses as large as $\approx 600$~MeV, see Ref.~\cite{Mai:2018djl} for further discussions.

With the 2-body input fixed, the only remaining unknown of the 3-body quantization condition remains the genuine 3-body force term $C(q,p;W)$. The functional form of this term is not known. We found, however, that the simplest choice $C(q,p,W)=c\,\delta^{(3)}(\boldsymbol{p}-\boldsymbol{q})$ leads to a good fit to the LQCD data~\cite{Beane:2007es,Detmold:2008fn} ($\chi^2_{\rm dof}=0.05$ for $c=0.2\pm1.5\cdot 10^{-10}$). The value of constant $c$ turns out to be of the same order of magnitude as the $\eta_L^3$ term introduced on the level of Hamiltonian in the large-volume expansion formula~\cite{Detmold:2008gh}. The comparison with the data as well as prediction of the excited levels for the $\pi^+\pi^+\pi^+$ system is depicted in the right panel of Fig.~\ref{pic:2body}. While no uncertainty bands (from the 2-body and 3-body input) are depicted there, they are discussed in Ref.~\cite{Mai:2018djl}.

In conclusion, we have analyzed the finite-volume spectrum for the $\pi^+\pi^+$ and $\pi^+\pi^+\pi^+$ systems using experimental data and a non-perturbative ansatz for the 2-body amplitude. The $\pi^+\pi^+$ energy levels in finite volume have been predicted and agree nicely with the available lattice data. Using this input and fitting the genuine 3-body contact term to the threshold level determined by the NPLQCD collaboration we have predicted the finite volume spectrum of the $\pi^+\pi^+\pi^+$ system up to $W=5\,m_\pi$. This is the first prediction of excited levels in a physical 3-body system. The extensions of this approach to multi-channel systems and systems with higher spin is work in progress.



\begin{thebibliography}{99}
\bibitem{Bruns:2010sv} 
  P.~C.~Bruns, M.~Mai and U.~G.~Mei\ss ner,
    Phys.\ Lett.\ B {\bf 697}, 254 (2011)
    [arXiv:1012.2233 [nucl-th]].

\bibitem{AlvarezRuso:2010xr} 
  L.~Alvarez-Ruso,
    (Bled Workshops in Physics. Vol. 11 No. 1)
  [arXiv:1011.0609 [nucl-th]].

\bibitem{Lang:2014tia} 
  C.~B.~Lang, L.~Leskovec, D.~Mohler and S.~Prelovsek,
    JHEP {\bf 1404}, 162 (2014)
    [arXiv:1401.2088 [hep-lat]].

\bibitem{Lang:2016hnn} 
  C.~B.~Lang, L.~Leskovec, M.~Padmanath and S.~Prelovsek,
    Phys.\ Rev.\ D {\bf 95}, no. 1, 014510 (2017)
    [arXiv:1610.01422 [hep-lat]].

\bibitem{Kiratidis:2016hda} 
  A.~L.~Kiratidis, W.~Kamleh, D.~B.~Leinweber, Z.~W.~Liu, F.~M.~Stokes and A.~W.~Thomas,
    Phys.\ Rev.\ D {\bf 95}, no. 7, 074507 (2017)
    [arXiv:1608.03051 [hep-lat]].

\bibitem{Woss:2018irj} 
  A.~Woss, C.~E.~Thomas, J.~J.~Dudek, R.~G.~Edwards and D.~J.~Wilson,
    JHEP {\bf 1807}, 043 (2018)
    [arXiv:1802.05580 [hep-lat]].

\bibitem{Beane:2007es} 
  S.~R.~Beane, W.~Detmold, T.~C.~Luu, K.~Orginos, M.~J.~Savage and A.~Torok,
    Phys.\ Rev.\ Lett.\  {\bf 100}, 082004 (2008)
    [arXiv:0710.1827 [hep-lat]].

\bibitem{Detmold:2008fn} 
  W.~Detmold, M.~J.~Savage, A.~Torok, S.~R.~Beane, T.~C.~Luu, K.~Orginos and A.~Parreno,
    Phys.\ Rev.\ D {\bf 78}, 014507 (2008)
    [arXiv:0803.2728 [hep-lat]].
      
\bibitem{Luscher:1986pf} 
  M.~Luscher,
    Commun.\ Math.\ Phys.\  {\bf 105}, 153 (1986).

\bibitem{Luscher:1990ux} 
  M.~Luscher,
    Nucl.\ Phys.\ B {\bf 354}, 531 (1991).
        
\bibitem{Hammer:2017kms} 
  H.-W.~Hammer, J.-Y.~Pang and A.~Rusetsky,
    JHEP {\bf 1710}, 115 (2017)
    [arXiv:1707.02176 [hep-lat]].
      
\bibitem{Hammer:2017uqm} 
  H.~W.~Hammer, J.~Y.~Pang and A.~Rusetsky,
    JHEP {\bf 1709}, 109 (2017)
    [arXiv:1706.07700 [hep-lat]].
      
\bibitem{Briceno:2017tce} 
  R.~A.~Briceño, M.~T.~Hansen and S.~R.~Sharpe,
    Phys.\ Rev.\ D {\bf 95}, no. 7, 074510 (2017)
    [arXiv:1701.07465 [hep-lat]].
      
\bibitem{Doring:2018xxx} 
  M.~Döring, H.~W.~Hammer, M.~Mai, J.-Y.~Pang, A.~Rusetsky and J.~Wu,
    Phys.\ Rev.\ D {\bf 97}, no. 11, 114508 (2018)
    [arXiv:1802.03362 [hep-lat]].
      
\bibitem{Sharpe:2017jej} 
  S.~R.~Sharpe,
    Phys.\ Rev.\ D {\bf 96}, no. 5, 054515 (2017)
    [arXiv:1707.04279 [hep-lat]].
    
\bibitem{Guo:2016fgl} 
  P.~Guo,
    Phys.\ Rev.\ D {\bf 95}, no. 5, 054508 (2017)
    [arXiv:1607.03184 [hep-lat]].
      
\bibitem{Hansen:2016ync} 
  M.~T.~Hansen and S.~R.~Sharpe,
    Phys.\ Rev.\ D {\bf 95}, no. 3, 034501 (2017)
    [arXiv:1609.04317 [hep-lat]].
      
\bibitem{Hansen:2016fzj} 
  M.~T.~Hansen and S.~R.~Sharpe,
    Phys.\ Rev.\ D {\bf 93}, no. 9, 096006 (2016)
  Erratum: [Phys.\ Rev.\ D {\bf 96}, no. 3, 039901 (2017)]
    [arXiv:1602.00324 [hep-lat]].
      
\bibitem{Hansen:2015zga} 
  M.~T.~Hansen and S.~R.~Sharpe,
    Phys.\ Rev.\ D {\bf 92}, no. 11, 114509 (2015)
    [arXiv:1504.04248 [hep-lat]].
      
\bibitem{Meissner:2014dea} 
  U.~G.~Meißner, G.~Ríos and A.~Rusetsky,
    Phys.\ Rev.\ Lett.\  {\bf 114}, no. 9, 091602 (2015)
  Erratum: [Phys.\ Rev.\ Lett.\  {\bf 117}, no. 6, 069902 (2016)]
    [arXiv:1412.4969 [hep-lat]].
      
\bibitem{Briceno:2012rv} 
  R.~A.~Briceno and Z.~Davoudi,
    Phys.\ Rev.\ D {\bf 87}, no. 9, 094507 (2013)
    [arXiv:1212.3398 [hep-lat]].

\bibitem{Bour:2012hn} 
  S.~Bour, H.-W.~Hammer, D.~Lee and U.~G.~Mei\ss ner,
    Phys.\ Rev.\ C {\bf 86}, 034003 (2012)
    [arXiv:1206.1765 [nucl-th]].

\bibitem{Kreuzer:2012sr} 
  S.~Kreuzer and H.~W.~Grie\ss hammer,
    Eur.\ Phys.\ J.\ A {\bf 48}, 93 (2012)
    [arXiv:1205.0277 [nucl-th]].
      
\bibitem{Polejaeva:2012ut} 
  K.~Polejaeva and A.~Rusetsky,
    Eur.\ Phys.\ J.\ A {\bf 48}, 67 (2012)
    [arXiv:1203.1241 [hep-lat]].

\bibitem{Agadjanov:2016mao} 
  D.~Agadjanov, M.~Doring, M.~Mai, U.~G.~Meißner and A.~Rusetsky,
    JHEP {\bf 1606}, 043 (2016)
    [arXiv:1603.07205 [hep-lat]].

\bibitem{Hansen:2017mnd} 
  M.~T.~Hansen, H.~B.~Meyer and D.~Robaina,
    Phys.\ Rev.\ D {\bf 96}, no. 9, 094513 (2017)
    [arXiv:1704.08993 [hep-lat]].

\bibitem{Mai:2017vot} 
  M.~Mai, B.~Hu, M.~Doring, A.~Pilloni and A.~Szczepaniak,
    Eur.\ Phys.\ J.\ A {\bf 53}, no. 9, 177 (2017)
    [arXiv:1706.06118 [nucl-th]].

\bibitem{Mai:2017bge} 
  M.~Mai and M.~Döring,
    Eur.\ Phys.\ J.\ A {\bf 53}, no. 12, 240 (2017)
    [arXiv:1709.08222 [hep-lat]].

\bibitem{Mai:2018djl} 
  M.~Mai and M.~Doring,
    arXiv:1807.04746 [hep-lat].
  
\bibitem{Bedaque:1999vb} 
  P.~F.~Bedaque and H.~W.~Griesshammer,
    Nucl.\ Phys.\ A {\bf 671}, 357 (2000)
    [nucl-th/9907077].

\bibitem{Mai:2017wdv} 
  M.~Mai, B.~Hu, M.~Doring, A.~Pilloni and A.~Szczepaniak,
  PoS Hadron {\bf 2017}, 140 (2018).  

\bibitem{Gasser:1983yg} 
  J.~Gasser and H.~Leutwyler,
    Annals Phys.\  {\bf 158}, 142 (1984).
  
\bibitem{Truong:1988zp} 
  T.~N.~Truong,
    Phys.\ Rev.\ Lett.\  {\bf 61}, 2526 (1988).
        
\bibitem{Detmold:2008gh} 
  W.~Detmold and M.~J.~Savage,
  Phys.\ Rev.\ D {\bf 77}, 057502 (2008)
  [arXiv:0801.0763 [hep-lat]].


\end{thebibliography}
\end{document}